\documentclass[11pt,journal]{IEEEtran}
\usepackage{amsfonts,amsmath,amssymb,amsthm,epsfig}
\usepackage[mathcal]{euscript} 

\newcommand{\pr}{{\mathbb{P}}}
\newcommand{\ex}{{\mathbb{E}}}
\newcommand\cE{\mathcal{E}}
\newcommand{\openone}{\leavevmode\hbox{\small1\normalsize\kern-.33em1}}

\newtheorem{thm}{Theorem}

\newtheorem{cor}{Corollary}

\newtheorem{defs}{Definition}

\DeclareMathOperator{\poly}{poly}

\begin{document}

\title{Training-Based Schemes are Suboptimal for High Rate Asynchronous
Communication}

\author{Venkat Chandar, Aslan Tchamkerten, and Gregory W. Wornell
\thanks{This work was supported in part by the National Science
  Foundation under Grant No.~CCF-0635191, and by a University IR\&D
  Grant from Draper Laboratory.}
\thanks{V.~Chandar and G.~W.~Wornell
  are with the Dept. EECS, Massachusetts Institute of Technology. (Email:
  \{vchandar,gww\}@mit.edu).  A.~Tchamkerten is with Telecom
  ParisTech, COMELEC. (Email: aslan.tchamkerten@telecom-paristech.fr).}}

\maketitle

\begin{abstract} 
We consider asynchronous point-to-point communication. Building on a
recently developed model, we show that training based schemes, i.e.,
communication strategies that separate synchronization from
information transmission, perform suboptimally at high rate.
\end{abstract}


\section{ Model and Review of Results}
\label{pform}

We consider the asynchronous communication setting developed in \cite{TCW},
which provides an extension to
Shannon's original point-to-point model for synchronous communication~\cite{Sha2}. 

We recall the setting in \cite{TCW}. Communication takes place over a discrete
memoryless channel characterized by its finite input and output alphabets
$\mathcal{X}$ and $\mathcal{Y}$, respectively, and transition probability
matrix $Q(y|x)$, for all $y\in {\mathcal{Y}}$ and $x\in {\mathcal{X}}$.  There
are $M\geq 2$ messages $\{1,2,\ldots,M\}$. For each message $m$ there is an
associated codeword $c^N(m)\triangleq c_1(m)c_2(m)\ldots c_N(m)$, a string of
length $N$ composed of symbols from $\mathcal{X}$.\footnote{The symbol `$\triangleq$'
stands for `equal by definition.'} The $M$ codewords form a codebook
${\mathcal{C}}_N$.  The transmitter
selects a message $m$, randomly and uniformly over the message set, and starts
sending the corresponding codeword $c^N(m)$ at a random time $\nu$, unknown to
the receiver, independent of $c^N(m)$, and uniformly distributed in
$\{1,2,\ldots,A\}$. The transmitter and the receiver know the integer $A\geq
1$, which we refer to as the \emph{asynchronism level} between the transmitter
and the receiver.  If $A=1$ the channel
is said to be synchronized. The capacity of the synchronized channel $Q$ is
denoted $C$, or $C(Q)$ when necessary for clarity.

During information transmission the receiver observes a noisy
version of the sent codeword, while before and after the information
transmission it observes only noise. Conditioned on the event $\{\nu=k\}$, $k\in
\{1,2,\ldots,A\}$,
and on the message $m$ to be conveyed, the receiver observes
independent symbols $Y_1,Y_2,\ldots$ distributed as follows. If $i\in
\{1,2,\ldots,k-1\}$ or $i\in \{k+N,k+N+1,\ldots, A+N-1\}$, the
distribution of $Y_i$ is $$Q_\star(\cdot)\triangleq Q(\cdot|\star)$$ for some
fixed $\star \in \mathcal{X}$. At any time $i\in \{k ,k+1,\ldots,
k+N-1\}$, the distribution of $Y_i$ is $$Q(\cdot|{c_{i-k+1}(m)})\;.$$ It
should be emphasized that the transition probability matrix
$Q(\cdot|\cdot)$, together with the `no-input' symbol $\star$,
characterizes the communication channel.  In particular, the $\star$
is not a parameter of the transmitter, i.e., the system designer
cannot designate which symbol in the input alphabet is $\star$.
This symbol can, however, be used for the codebook design.
Throughout the paper, whenever we refer to a certain channel $Q$,
we implicitly assume that the $\star$ symbol is given.

 The decoder consists of a sequential test $(\tau_N,\phi_N)$,
where $\tau_N$ is a stopping time --- bounded by $A+N-1$ --- with respect to the output sequence
$Y_1,Y_2,\ldots$ indicating when decoding happens, and where $\phi_N$ denotes a
decision rule that declares the decoded message.
Recall that a stopping time $\tau$ (deterministic or randomized) is an
integer-valued random variable with respect to a sequence of random variables
$\{Y_i\}_{i=1}^\infty$ so that the event $\{\tau=n\}$, conditioned on
the realizations of $\{Y_i\}_{i=1}^{n}$,  is independent of those of
$\{Y_{i}\}_{i=n+1}^{\infty}$, for all
$n\geq 1$. The function $\phi_N$ is then defined as any
${\mathcal{F}}_{\tau_N}$-measurable map taking values in $\{1,2,\ldots,M\}$, where ${\mathcal{F}}_1,{\mathcal{F}}_2,\ldots$ is the
natural filtration induced by the output process $Y_1,Y_2,\ldots$. 

We are interested in {\emph{reliable and quick decoding}}. To that aim we first
define the average decoding error probability (given a codebook and a decoder)
as \begin{align*}
\pr(\cE)\triangleq\frac{1}{A}\frac{1}{M}\sum_{m=1}^M\sum_{k=1}^A
\pr_{m,k} (\cE), \end{align*} where $\cE$ indicates the
event that the decoded message does not correspond to the sent message, and
where the subscripts `$_{m,k}$' indicate the conditioning on the event that message $m$
starts being sent at time $k$.

Second, we define the average communication rate with respect to
the average delay it takes the receiver to react to a sent message, i.e.\footnote{$\ln $ denotes the natural logarithm. }
\begin{align*}
 R\triangleq \frac{\ln M}{\ex(\tau_N-\nu)^+}\triangleq \frac{\ln |{\mathcal{C}}_N|}{\ex(\tau_N-\nu)^+}
\end{align*}
 where $\ex(\tau_N-\nu)^+$ is defined as \begin{align*}
\ex(\tau_N-\nu)^+\triangleq\frac{1}{A}\frac{1}{M}\sum_{m=1}^M\sum_{k=1}^A
\ex_{m,k}( \tau_N-k)^+, \end{align*} where $\ex_{m,k}$ denotes the expectation with
respect to $\pr_{m,k}$, and where $x^+$ denotes
$\max\{0,x\}$. With the above definitions, we now recall
the notions of $(R,\alpha)$ coding scheme and capacity function.
\begin{defs}[$(R,\alpha)$ coding scheme]\label{tasso}
Given a channel $Q$, a pair $(R,\alpha)$ is {\emph{achievable}} if there exists a
sequence $\{({\mathcal{C}}_N,(\tau_N,\phi_N)\}_{N\geq 1}$ of codebook/decoder
pairs that asymptotically achieves a rate $R$ at an asynchronism
exponent $\alpha$. This means that, for any $\varepsilon>0$ and all $N$ large
enough, the pair $({\mathcal{C}}_N,(\tau_N,\phi_N))$
\begin{itemize}
\item
operates under asynchronism level  $A=e^{(\alpha - \varepsilon)
N}$;
\item
 yields an average rate at least equal to $R-\varepsilon$; 
\item
achieves an average error
probability $\pr(\cE)$ at most equal to $\varepsilon$.
\end{itemize}
Given a channel $Q$, {\emph{an $(R,\alpha)$ coding scheme}} is a
sequence $\{({\mathcal{C}}_N,(\tau_N,\phi_N))\}_{N\geq 1}$ that achieves a rate $R$ at an asynchronism
exponent $\alpha$ as $N\rightarrow \infty$.
\end{defs}
\begin{defs}[Capacity of an asynchronous discrete memoryless channel]\label{caps}
 The capacity of an asynchronous discrete memoryless channel with (synchronized)
 capacity $C(Q)$ is the function
 \begin{align*}
 [0,C(Q)]&\rightarrow {\mathbb{R}}_+\\
 R&\mapsto \alpha(R,Q),
 \end{align*} 
 where $\alpha(R,Q)$ is the supremum of the set of asynchronism exponents that are
achievable at rate~$R$. 
\end{defs}
It turns out that the exponential scaling of the asynchronism exponent with respect to the
codeword length in Definition~\ref{tasso} is natural: asynchronism induces a rate loss with respect to the capacity of the
synchronous channel only when it grows at least exponentially with the
codeword length~\cite{TCW}.

The following theorem, given in  \cite{TCW2}, provides a non-trivial lower bound
to the capacity of asynchronous channels:
 \begin{thm}\label{ach}
For a given channel $Q $, let $\alpha \geq 0$ and let $P$ be a distribution over $\mathcal{X}$ such
that
\begin{equation*} 
\min_{V}\max\{D(V\|(PQ)_\mathcal{Y}),D(V\|Q_\star)\}>\alpha
\end{equation*}
where the minimization is over all distributions over $\cal{Y}$,
and where the distribution $(PQ)_\mathcal{Y}$ is defined as
$(PQ)_\mathcal{Y}(y)=\sum_{x\in {\mathcal{X}}}P(x)Q(y|x)$, $y\in
\mathcal{Y}$. 
Then, the pair $(R=I(PQ),\alpha)$
is achievable.
\end{thm}
\begin{cor}
At capacity, it is possible to achieve a strictly positive
asynchronism exponent, except for the case
when $Q_\star$ corresponds to the capacity-achieving output
distribution of the synchronous channel.\footnote{To see this, recall that, given the
channel $Q$, all
capacity-achieving input distributions $P$ induce the same output
distribution $(PQ)_\mathcal{Y}$. Whenever $(PQ)_\mathcal{Y}$ differs from
$Q_\star$, the min-max expression in Theorem~\ref{ach} is strictly positive.
Therefore capacity is achievable at a strictly positive asynchronism exponent.}
Moreover, the asynchronism exponent achievable at capacity can be arbitrarily
large, depending on the channel. 
\end{cor}
This is in contrast with training-based
schemes. The contribution of this paper, given in the next
section,  is to show that training-based scheme, in general, achieve a vanishing
asynchronism exponent in the limit of the rate going to capacity.

\section{Training-Based Schemes}

The usual approach to communication is a training-based architecture.
In such schemes, each codeword is composed of two parts. The first
part, the sync preamble, is a sequence of symbols common to all
the codewords, hence carries no information;  its only purpose is
to help the decoder to locate the
sent message. The second part carries information. The decoder
operates according to a two-step procedure. First it tries to locate
the codeword by seeking the sync preamble. Once the sync preamble is
located, it declares a message based on the subsequent symbols. A
formal definition of a training-based scheme follows.
\begin{defs}
A training-based scheme is a coding scheme $\{({\mathcal{C}}_N,(\tau_N,\phi_N))\}_{N\geq 1}$
with  the following properties. For some $\varepsilon>0$, $\eta\in [0,1]$, and all
integers $N\geq 1$ 
\begin{itemize}
\item[i.]
each codeword in ${\mathcal{C}}_N$ starts with a string of size $\eta N$ that is
common to all codewords;\footnote{To be precise, the string size should be an
integer, and instead of having it equal to $\eta N$ we should have it equal to 
$\left\lfloor \eta N \right\rfloor$. However, since we are interested in the
asymptotic $N\rightarrow \infty$, this discrepancy typically
vanishes. Similar discrepancies are ignored throughout the paper.}
\item[ii.]
the decision time $\tau_N$ is such that the event
$\{\tau_N=n\}$, conditioned on the $\eta N$ observations $Y_{n- N+1}^{n-N+\eta
N}$,\footnote{We use $Y_i^j$ for $Y_i,Y_{i+1},\ldots,Y_j$ (for $i\leq j$).} is independent of all  other past observations, i.e.,
$Y_1^{n- N}$ and $Y_{n-N+\eta N+1}^{n}$;
\item[iii.]  the codebook ${\mathcal{C}}_N$ and the decoding time $\tau_N$ satisfy
$$\pr(\tau_N\geq k+2 N-1|\tau_N\geq k + N, \nu=k)\geq \varepsilon $$
for all $k\in \{1,2,\ldots,A\}$\;.
\end{itemize}
\end{defs}
Condition i. specifies the size of the sync preamble. Condition ii. indicates that the decoding
time should depend only on the sync preamble. Condition iii. imposes that the codeword symbols that follow
the sync preamble should not be used to help the decoder locate the codeword.
If we remove Condition iii.,  one
could imagine having information symbols with a `sufficiently biased' distribution to help the
decoder locate the codeword position (the `information symbols' could even start
with a second preamble!). In this case the sync preamble is followed by
a block of information symbols that also helps the decoder to locate the
sent codeword. To avoid this, we impose Condition iii. which says that, once the sync preamble is
missed (this is captured by the event $\{\tau_N\geq k+N, \nu=k\}$, the
decoder's decision to stop will likely no more depend on the sent codeword since it
will occur after $ k+2N-1$. 

Finally, it can be shown that
a large class of  training-based schemes considered in practice satisfy the
above three conditions.
\begin{thm}\label{firsttraining} 
A training-based scheme that achieves a rate $R\in (0,C(Q)]$ operates at an
asynchronism exponent $\alpha$ upper bounded as 
\begin{equation*} 
\alpha\leq \left(1-\frac{R}{C}\right)\max_{P}\min_{W}\max\{D_1,D_2\},
\end{equation*}
where $D_1\triangleq D(W\|Q|P)$, and $D_2\triangleq
D(W\|Q_\star|P)$.\footnote{We use the standard notation $D(W\|Q|P)$ for the
Kullback-Leibler distance between the joint distributions
$P(\cdot)W(\cdot|\cdot)$ and $P(\cdot)Q(\cdot|\cdot)$ (see, e.g., \cite[p. 31]{CK}).}  The first maximization is over all distributions over
$\mathcal{X}$ and the minimization is over all conditional
distributions defined over ${\mathcal{X}}\times {\mathcal{Y}}$.
\end{thm} 

The following result is a consequence of Theorem~\ref{firsttraining}.
\begin{cor} \label{coratcap}
Unless the no-input symbol $\star$ does not generate a
particular channel output symbol (i.e., $Q(y|\star)=0$ for some $y\in
{\mathcal{Y}}$), training-based schemes achieve a vanishing asynchronism
exponent as $R\rightarrow C(Q)$.
\end{cor}
\begin{IEEEproof}[Proof of Corollary~\ref{coratcap}]
We consider the inequality of Theorem~\ref{firsttraining} and first upper bound
the minimization by choosing $W=Q$. With this choice, the inner
maximization becomes $D_2=D(Q||Q_\star|P)$ (since $D_1=D(Q||Q|P)=0$). Maximizing
over $P$ yields
$$\max_P D(Q||Q_\star|P)= \max_{x\in \mathcal{X}}D(Q(\cdot|x)||Q_\star)$$
which is bounded when $Q(y|\star)>0$ for all $y\in \mathcal{Y}$. Therefore the
max-min-max term in the inequality of Theorem~\ref{firsttraining} is finite and gets multiplied by a term that vanishes as
$R\rightarrow C(Q)$.
\end{IEEEproof}
Thus, except for degenerate cases, training-based schemes achieve a
vanishing asynchronism exponent in the limit of the rate going to
capacity. In contrast, from Theorem \ref{ach} one deduces that it is
possible, in general, to achieve a non-zero asynchronism exponent at
capacity, as we saw above.

This suggests that to achieve a high rate under strong
asynchronism, separating synchronization
from information transmission is suboptimal;  the codeword
symbols should all play the dual role of  information carriers and
`information flags.' 

\subsection*{Sketch of Proof of Theorem~\ref{firsttraining}}

Consider a training-based scheme
$\{({\mathcal{C}}_N,(\tau_N,\phi_N))\}_{N\geq 1}$. For simplicity, we assume
that the sync preamble distribution of ${\mathcal{C}}_N$ is the same, equal to
$P$, for all $N\geq 1$. The case of different preamble distributions for different
values of $N$ requires a minor extension. The proof consists in showing that if the following two inequalities hold
\begin{align}
\eta D(W||Q|P)&< \alpha \label{p1}\\
\eta D(W||Q_\star|P)&< \alpha \label{p2}
\end{align}
for some conditional distribution $W$, then the average reaction delay achieved
by $\{({\mathcal{C}}_N,(\tau_N,\phi_N))\}_{N\geq 1}$ grows exponentially with $N$. This, in turn, can
be shown to imply that the rate is asymptotically equal to zero. Therefore,
maximizing over the sync preamble distributions, it is necessary that 
$$\alpha \leq \eta \max_P\min_W\max\{D(W||Q|P),D(W||Q_\star|P)\}$$
in order to achieve a strictly positive rate $R$.
The second part of the proof, omitted in this paper, consists in showing that the highest value of $\eta$ compatible with rate $R$
communication is upper bounded by $(1-R/C(Q))$. This with
the above inequality yields the desired result.

Below we sketch the argument that shows that, if both \eqref{p1} and \eqref{p2}
hold, the average reaction delay grows exponentially with $N$.

To keep the presentation simple,  in the equations below we omit
terms that go to zero in the limit $N\rightarrow \infty$.
Thus, although the equations may not be valid as written, they become valid
in that limit.

Let $\{({\mathcal{C}}_N,(\tau_N,\phi_N))\}_{N\geq 1}$ be a training-based scheme
with preamble empirical distribution equal to $P$.  By property
ii.,
the stopping time $\tau_N$ is such that the event $\{\tau_N=n\}$
depends only on the realizations of $Y_{n- N+1}^{n-N+\eta N}$.
For simplicity, instead of $\tau_N$, we are going to consider the shifted
stopping time $\tau'_N\triangleq \tau_N-(1-\eta)N$ whose
decision to stop at a certain moment depends on immediate $\eta N$ previously observed
symbols. Clearly,  $\tau'_N$ can be written as
$$\tau'_N=\inf\{i\geq 1: S_i=1\},$$
where each $S_i$ is some  (decision) function defined over $Y_{i-\eta N+1}^i$ and that
take on the values $0$ or $1$. 

The condition iii. in terms of $\tau'_N$ becomes
\begin{align}\label{prop3mod}
\pr(\tau_N'\geq k+N+\eta N-1|\tau_N'\geq k +\eta N, \nu=k)\geq \varepsilon 
\end{align}
for all $k\in \{1,2,\ldots,A\}$.

Let us define the events
\begin{align*}
\cE_1&=\{\tau_N'\geq \nu+\eta N\}\\
\cE_2&=\{S_i=0 \:\: \text{for}\:\:  i\in\{\nu+N+\eta N-1,\ldots,3A/4\}\}\\
\cE_3&=\{\tau_N'\geq \nu+N+\eta N-1\}\\
\cE_4&=\{\nu\leq A/4\}\;.
\end{align*}
We lower bound the reaction delay as
\begin{align}\label{un}
\ex((\tau_N'-\nu)^+)\geq \ex((\tau_N'-\nu)^+|\cE_1,\cE_4)\pr (\cE_1,\cE_4),
\end{align}
and consider the two terms on the right-side separately. 

We first show that  $\ex((\tau_N'-\nu)^+|\cE_1,\cE_4)=
\Omega(A)$.\footnote{$\Omega(\cdot)$ refers to the standard
Landau order notation.}
We have 
\begin{align}\label{indeeduseful}
\ex&((\tau_N'-\nu)^+|\cE_1,\cE_4)\nonumber \\
&\geq \ex((\tau_N'-\nu)^+|\cE_1,\cE_2,\cE_3,\cE_4)\pr(\cE_2,\cE_3|\cE_1,\cE_4)\nonumber \\
&=\ex((\tau_N'-\nu)^+|\cE_2,\cE_3,\cE_4)\pr(\cE_2,\cE_3|\cE_1,\cE_4)\nonumber \\
&=\ex((\tau_N'-\nu)^+|\tau'_N\geq 3A/4, \nu\leq A/4)\pr(\cE_2,\cE_3|\cE_1,\cE_4)\nonumber \\
&\geq \frac{A}{2}\pr(\cE_2,\cE_3|\cE_1,\cE_4)
\end{align}
where the first equality holds since
$\cE_3\subset\cE_1$, and where the second equality
holds since $\cE_2\cap \cE_3=\{\tau'_N>3A/4\}$.
We now prove that  $\pr(\cE_2|\cE_1,\cE_4)$ and
$\pr(\cE_3|\cE_1,\cE_4)$ have large probabilities for large $N$. This  implies that 
$\pr(\cE_2,\cE_3|\cE_1,\cE_4)$ has a probability bounded away from zero for $N$
large enough. This together with \eqref{indeeduseful} implies that
$\ex((\tau_N'-\nu)^+|\cE_1,\cE_4)=\Omega(A)$ as claimed above.

For $\pr(\cE_2|\cE_1,\cE_4)$ we have
\begin{align}\label{e21}
\pr(\cE_2|\cE_1,\cE_4)&=\pr(\cE_2|\cE_4)\nonumber \\
&= \pr(S_{\nu+N+\eta N-1}^{A/4}=0|\nu\leq A/4)\nonumber \\
&=\frac{1}{A/4} \sum_{k=1}^{A/4}\pr(S_{k+N+\eta N-1}^{A/4}=0|\nu=k)\nonumber \\
&=\frac{1}{A/4} \sum_{k=1}^{A/4}\pr_\star(S_{k+N+\eta N-1}^{A/4}=0)\nonumber \\
&\geq\frac{1}{A/4} \sum_{k=1}^{A/4}\pr_\star(S_{1}^{A/4}=0)\nonumber \\
&= \pr_\star(S_{1}^{A/4}=0)\nonumber \\
&=  \pr_\star(\tau'_N>3A/4)
\end{align}
where $\pr_\star$ denotes the output distribution under pure noise,
i.e., when the $Y_i$'s are i.i.d. according to $Q_\star$.
For the first equality we used the independence between $\cE_2$
and $\cE_1$ conditioned on $\cE_4$. For the fourth equality we
noted that, conditioned on $\{\nu=k\}$, the event
$S_{k+N+\eta N-1}^{3A/4}$ is
independent of the sent codeword (prefix and information sequence),
hence its probability is $\pr_\star$.

 Now,
the event $\{\tau'_N>3A/4\}$ only depends on the output symbols up to time $3A/4$. The probability
of this event under $\pr_\star$ is thus the same as under the probability distribution induced by
the sending of a message {\emph{after}} time $3A/4$. Therefore,
since the probability of error vanishes for large $N$, and that a
message starts being sent after time $3A/4$ with (large) probability
$1/4$,  we must have $\pr_\star(\tau'_N>3A/4)\approx 1$
for large $N$. Hence from \eqref{e21} we have
\begin{align}\label{e22}
\pr(\cE_2|\cE_1,\cE_4)\approx 1
\end{align}
for large $N$. Now consider $\pr(\cE_3|\cE_1,\cE_4)$. Using \eqref{prop3mod}, we have
\begin{align}\label{e34}
\pr(\cE_3|\cE_1,\cE_4)\geq \varepsilon.
\end{align}
From \eqref{e22} and \eqref{e34} we deduce that
$\pr(\cE_2,\cE_3|\cE_1,\cE_4)$ is the (conditional) probability of
the intersection of two large probability events. Therefore
$\pr(\cE_2,\cE_3|\cE_1,\cE_4)$ has a
probability bounded away from zero as $N\rightarrow \infty$. Hence, we have shown
that
\begin{align}\label{deux}
\ex((\tau_N'-\nu)^+|\cE_1,\cE_4) = \Omega(A)
\end{align}
as claimed earlier.

Second, we prove that 
\begin{align}\label{tofp}
\pr(\cE_1,\cE_4)=\Omega(e^{-\eta N D_1}\text{poly}(N)),
\end{align}
where $D_1=D(W\|Q|P)$, $P$ denotes the type of the preamble, and $\poly(N)$
denotes a quantity that goes to $0$ at most polynomially quickly
as a function of $N$.
 
We expand $\pr(\cE_1,\cE_4)$ as
\begin{align}\label{ez1}
\pr(\cE_1,\cE_4)=\frac{1}{A}\sum_{k=1}^{A/4}\pr_k(\tau'_N\geq k+\eta N),
\end{align}
where $\pr_k$ represents the probability distribution of the output conditioned on the 
event $\{\nu = k\}$.
Further, by picking a conditional distribution $W$ defined over ${\cal{X}}\times{\cal{Y}}$
such that $\pr_k(Y_k^{k+\eta N-1}\in {\mathcal{T}}_W^{\eta
N}(P))>0$,\footnote{The set ${\mathcal{T}}_W^{\eta
N}(P)$ corresponds to all output sequences $y^{\eta N}$ that,  together with the
preamble, have joint type equal to $P(\cdot)W(\cdot|\cdot)$. } we
lower the term in the above sum as
\begin{align}\label{ez2}
\nonumber \pr_k(\tau'_N\geq k+\eta N)\geq &\pr_k(\tau'_N\geq k+\eta N|Y_k^{k+\eta N-1}\in {\mathcal{T}}_W^{\eta
N}(P))\\
&\hspace{.3cm}\times \pr_k (Y_k^{k+\eta N-1}\in {\mathcal{T}}_W^{\eta
N}(P))\;.
\end{align}
We lower bound each of the two terms on the right-side of~\eqref{ez2}.

For the first term,
a change of measure argument reveals that 
\begin{align}\label{ez4}\pr_k&(\tau'_N\geq k+\eta N|Y_k^{k+\eta N-1}\in {\mathcal{T}}_W^{\eta
N}(P))\nonumber \\
&=\pr_\star(\tau'_N\geq k+\eta N|Y_k^{k+\eta N-1}\in {\mathcal{T}}_W^{\eta
N}(P))\;.
\end{align}
To see this, one expands  $$\pr_k(\tau'_N\geq k+\eta N|Y_k^{i+\eta N-1}\in
{\mathcal{T}}_W^{\eta N}(P))$$ by further conditioning on individual sequences in ${\mathcal{T}}_W^{\eta
N}(P)$. Then, one uses the fact that, conditioned on a particular such sequence, 
the channel outputs outside the time window $\{k,k+1,\ldots,k+\eta N-1\}$ are
distributed according to noise, i.e., i.i.d. according
to~$Q_\star$. 

For the second term we have
\begin{align}\label{ez3}
\pr_k(Y_k^{k+\eta N-1} \in {\mathcal{T}}_W^{\eta
N}(P) )\geq \text{poly}(N)e^{-\eta N D_1}
\end{align}
using \cite[Lemma 2.6, p. 32]{CK}, where $D_1\triangleq D(W\|Q|P)$.
  Combining \eqref{ez1}, \eqref{ez2}, \eqref{ez4}, and \eqref{ez3} we get 
\begin{align}\label{ez5}
\pr&(\cE_1,\cE_4)&\nonumber \\
\geq &\text{poly}(N)\frac{e^{-\eta N D_1}}{A}\times \nonumber \\
&\times\sum_{k=1}^{A/4}\pr_\star(\tau'_N\geq
i+\eta N|Y_k^{k+\eta N-1}\in {\mathcal{T}}_W^{\eta N}(P))\nonumber \\
\geq &\text{poly}(N)\frac{e^{-\eta N (D_1-D_2)}}{A}\times \nonumber \\
&\times \sum_{k=1}^{A/4}\pr_\star(\tau'_N\geq
i+\eta N, Y_k^{i+\eta N-1}\in {\mathcal{T}}_W^{\eta N}(P))\;,
\end{align}
where $D_2 \triangleq D(W\|Q_\star|P)$, and where for the second
inequality we again used  \cite[Lemma 2.6, p. 32]{CK}.

Now,  assuming that $\alpha > \eta D_2$,
one can show that 
$$\sum_{k=1}^{A/4}\pr_\star(\tau'_N\geq
k+\eta N,Y_k^{k+\eta N-1}\in {\mathcal{T}}_W^{\eta N}(P))=\Omega (Ae^{-\eta D_2})$$
using the union bound. Therefore, under the above assumption we get from
\eqref{ez5} the desired claim that
\begin{align}\label{ez6}
\pr&(\cE_1,\cE_4)=\Omega(e^{-\eta N D_1}\text{poly}(N))\,.
\end{align}
From \eqref{un}, \eqref{deux}, and \eqref{ez6}, we conclude that if $\alpha > \eta D_2$
then
$$\ex((\tau_N'-\nu)^+)\geq \Omega (Ae^{-\eta N D_1}\text{poly}(N))\;.$$
Therefore, letting $A=e^{N\alpha}$, we deduce that, if, in addition to the
inequality $\alpha > \eta D_2$, we also have $\alpha > \eta D_1$, the average
reaction delay $\ex((\tau_N'-\nu)^+)$ grows exponentially with $N$. 
\hfill\IEEEQEDclosed

\section*{Concluding Remarks} Synchronization and information
transmission of virtually all practical communication systems are performed
separately, on the basis of different communication bits. Moreover, in general,
the rate of these strategies is computed with respect to the
information transmission time period, ignoring the delay overhead caused by various
hand-shake protocols used to guarantee synchronization. In these cases, the
notions of `high rate' or `capacity-achieving' communication strategies clearly
raises questions. 

Building on an extension of Shannon's original point-to-point synchronous
communication channel model to assess the overall rate performance of
asynchronous communication systems, we showed that training-based schemes
perform suboptimally at high rates. In this regime, it is necessary to envision
communication strategies that integrate synchronization into information
transmission.

\section*{Acknowledgments}
We thank the reviewer for valuable comments.

\bibliographystyle{IEEEtran}
\bibliography{../../../common_files/bibiog}
\end{document}